\documentstyle[preprint,prl,aps]{revtex}
\begin{document}
\draft

\title{The effect of the Coulomb interaction on the mesoscopic
persistent current}

\author{D. Yoshioka and H. Kato}
\address{Institute of Physics,
         College of Arts and Sciences,
         University of Tokyo, \\Komaba, Meguro-ku, Tokyo 153, Japan}
\date{\today}

\maketitle
\begin{abstract}
The persistent current in three-dimensional mesoscopic rings is
investigated numerically.
The model is tight-binding one with random site-energies and
interaction between electrons.
The self-consistent
Hartree-Fock approximation is adopted for the interaction between
electrons, and models with up to $6\times 6\times 20$ sites
are investigated.
It is shown that the long-range Coulomb interaction enhances the
persistent current  for rings with finite width.
The origin of the enhancement is discussed.
\end{abstract}
\vskip 0.5truecm
\leftline{Keywords: persistent current, mesoscopic ring, Coulomb interaction}
\vskip 2.truecm
\leftline{Correspondence to: D. Yoshioka, Institute of Physics, University of
Tokyo }
\leftline{~~~~~~~~~~~~~~~~~~~~~~~~~~~~~Meguro-ku, Tokyo 153, Japan}
\leftline{~~~~~~~~~~~~~~~~~~~~~~~~~~~~~FAX: 81-3-5454-6526}
\leftline{~~~~~~~~~~~~~~~~~~~~~~~~~~~~~e-mail: daijiro@tansei.cc.u-tokyo.ac.jp}


\narrowtext

\section{Introduction}
\label{sec:intro}
It has been predicted by  B\"uttiker {\em et al.} \cite{but} that
an equilibrium persistent current flows
in a mesoscopic normal-metal ring threaded by a magnetic flux.
This prediction has been experimentally confirmed\cite{exp1,exp2,exp3}.
Especially, the experiment on single Au ring\cite{exp2}
revealed that the size of the
persistent current observed was quite large.
Namely the observed current was $I = (0.3 \sim 2.0) I_0$,
where $I_0 = ev_{F}/L$
is of the order of the persistent current in an ideal one-dimensional
ring with $v_F$ being the Fermi velocity and $L$ being the
circumference of the ring.
The current is periodic in the magnetic flux $\phi$ piercing the
ring with period
$\phi_0=h/e$ as expected.
Thus the persistent current was predicted theoretically and confirmed
by the experiment.
However, this is not the whole story.
Quantitative understanding of the current has not been established.

It has been clarified by numerical\cite{mon1,mon2}
and theoretical\cite{altland} investigations
that theory without taking into account the
interaction between electrons can give current which is too small
compared to the experiments\cite{exp1,exp2}.
Inclusion of the mutual interaction into theory is hard and difficult
to assess the validity of the approximation employed, but
several groups\cite{ae,schmid}
have done such calculations, and obtained results that
sample averaged current is of the order of $(l_e/L)I_0$,
 where $l_e$ is the elastic mean free path.
In this case the current is periodic in
$\phi_0/2$, because sample averaging kills the component with period
$\phi_0$.
This current is much smaller than $I_0$ in the experimental situation
of $L >> l_e$.
In these theories the Coulomb interaction acts mainly to maintain the
local charge neutrality of the system.
Another theory by Kopietz\cite{kop} takes into account
long-range Coulomb interaction
energy associated with the charge fluctuation.
This theory gives the average persistent current of the order of
$0.1I_0$ with period $\phi_0/2$.
This value is large enough to explain the experiments\cite{exp1}.
However, since the fluctuation should be suppressed by the Coulomb
interaction itself to maintain the approximate local charge
neutrality, it is questionable that such large value survives
improvement of the theory\cite{vign,altl}.

On the other hand, numerical investigation is free from approximations,
and has a possibility to  give a clear answer to the origin of the
large persistent current.
However, unfortunately rigorous treatments are limited to quite small
system sizes.
Thus until now only small one-dimensional systems have been
investigated with negative answer to the role of the Coulomb
interaction\cite{ab,monc,gog}.
Namely, it has been clarified that in one-dimensional systems the
repulsive interaction between electrons suppress the persistent
current in situations relevant to the experiment.
We have also investigated the one-dimensional system with the Coulomb
interaction between electrons\cite{ky}.
Our results were the same as others.
However, we also found that the Hartree-Fock approximation
gives almost the same
results as those of the exact diagonalization.
The merit of the Hartree-Fock approximation is that we
can deal with much larger systems.
Thus we now apply our method to three-dimensional rings to
investigate the effect of the Coulomb interaction.
In the present work we have found that for the three-dimensional ring
the Coulomb interaction enhances the persistent current.

\section{Model}
\label{sec:model}

We consider the Anderson model with Coulomb interaction
between electrons.
For simplicity we neglect spin freedom of electrons.
Thus we consider a tight binding model on a cubic lattice.
Our model ring consists of
$ N_w \times N_h
 \times
N_\ell $
 sites:
circumference $N_\ell a$, width $N_w a$, and
height $N_ha$, where $a$ is the lattice constant.
The coordinate of the $i$-th site is given by three integers
$(\ell_i,w_i,h_i)$,
where $1 \leq \ell_i \leq N_\ell$, $1 \leq w_i \leq N_w$, and
$1 \leq h_i \leq N_h$.
Periodic boundary condition is imposed only in $\ell_i$,
namely $\ell_i +N_\ell = \ell_i$.
In the other two directions the lattice is truncated.
Each site has random site-energy $\varepsilon_i$, which
has uniform distribution between $-W/2$ and $W/2$.
Thus our Hamiltonian is
\begin{eqnarray}
\label{ex}
{\cal H}&=&-\sum_{i,j}( t_{i,j} e^{i \theta_{i,j}} c_i^{\dagger} c_j
        + h.c.)
        + \sum_i \varepsilon_i c_i^{\dagger} c_i \nonumber \\
    & & + \frac12 \sum_{i \ne j} V \frac{a}{r_{ij}}
       (c_i^\dagger c_i - \rho_0)(c_j^\dagger c_j - \rho_0),
\end{eqnarray}
where $c_i^\dagger (c_i)$ is the creation (destruction) operator of
a spinless electron,
$t_{i,j}$ is non-zero and equal to $t$ only for nearest neighbor
hopping, $\theta_{i,j}$ is non-zero only for hopping along the
circumference of the ring and has a value $2\pi\phi/N_\ell \phi_0$,
$V=e^2/4\pi\epsilon a$ is the strength of the Coulomb interaction,
and $\rho_0$ is neutralizing positive charge at each site.
To make a ring which has the same lattice constant everywhere we
have embedded
our three-dimensional ring in the four-dimensional space, so
the distance $r_{ij}$ between $i$, $j$ sites  is given by
\begin{eqnarray}
\label{dist}
(\frac{r_{ij}}{a})^2 &=& 2R^2\{1 - \cos[\frac{2\pi}{N_\ell}
(\ell_i - \ell_j)]\} \nonumber \\
& &+ (w_i - w_j)^2 + (h_i - h_j)^2,
\end{eqnarray}
where $R=[2\sin(\pi/N_\ell)]^{-1}$ is the radius of the ring.

The following treatment is the same as before\cite{ky}:
The interaction term is decoupled following standard Hartree-Fock
scheme by introducing the average $\langle c_i^\dagger c_j \rangle$,
which is determined self-consistently by iteration.
Once the ground state energy $E_{\rm g}(\phi)$ is determined
self-consistently, the persistent
current $I(\phi)$ is given by the derivative of the energy
$E_g(\phi)$ with respect to the magnetic flux:
\begin{eqnarray}
\label{pc}
I(\phi)=-\frac{\partial E_{\rm g}(\phi)}{\partial \phi},
\end{eqnarray}
Since $E_{\rm g}(\phi)$ is an even function of $\phi$ and periodic with
period $\phi_0$,
we first Fourier analyze the energy,
\begin{eqnarray}
\label{ef}
E_{\rm g}(\phi) = E_0+ \sum_{n=1}^\infty E_n
\cos(2\pi n \frac{\phi}{\phi_0}),
\end{eqnarray}
and obtain the Fourier components of the current:
\begin{eqnarray}
\label{cf}
I(\phi) &=& \sum_{n=1}^\infty \frac{2\pi n}{\phi_0} E_n
         \sin(2\pi n\frac{\phi}{\phi_0}) \nonumber \\
       &\equiv&\sum_{n=1}^{\infty} I_n \frac{t}{\phi_0}
        \sin(2\pi n\frac{\phi}{\phi_0}).
\end{eqnarray}
Here $I_n$ is a dimensionless number, which depends on realization of
the random site-energy and the strength of the Coulomb interaction.
In the following we mainly concentrate on the $\phi_0$-periodic
component, $I_1$.

\section{Results}
\label{sec:res}

We have investigated up to 720-site systems by this method, the
largest being $N_\ell = 20$, $N_w=6$, and $N_h=6$ with 300 electrons.
The randomness is chosen such that the sample is in the diffusive
regime,
which is realized around $W/t= 4$ for the samples we have considered.

For the largest sized system we have investigated five samples,
which mean five different realization of the random site energies.
The results for $I_1$ for these five samples are shown in Fig. 1,
where samples are distinguished by different symbols, circles, two kinds
of triangles (normal and upside down), squares, and diamonds.
The absolute values of $I_1$ are plotted,
the sign of $I_1$ is distinguished
by open and closed symbols, open meaning positive and closed meaning
negative value of $I_1$, respectively.
It can be seen that the current is enhanced in four samples out of five.
This is the first evidence that the Coulomb interaction can enhance the
persistent current.

On the other hand, it has been established that the Coulomb interaction
suppresses the persistent current for one-dimensional samples.
It is evident that the finite cross section of the sample is
responsible for the enhancement.
Thus to clarify the effect we have calculated persistent current
for systems with the same length but with smaller cross sections.
In such calculations we have chosen the number of electrons such that
it is about 40\% of the lattice sites.
The results are shown in Fig.2.
Here we show sample averaged typical current, which is defined as
$I_{\rm typ} = (\langle I_1^2 \rangle )^{1/2}$, where $\langle ~ \rangle$
means average over five realizations of the random site energies for each
sample size.
The system sizes considered are
 $1\times 1\times 20$ sites with 8 electrons,
 $2\times 2\times 20$ sites with 32 electrons,
 $3\times 3\times 20$ sites with 72 electrons,
 $4\times 4\times 20$ sites with 128 electrons,
 $5\times 5\times 20$ sites with 200 electrons, and
 $6\times 6\times 20$ sites with 300 electrons,
which are shown by open circles, closed circles, open triangles
closed triangles, open squares, and closed squares, respectively.
It is seen that the typical current in the absence of the interaction
increases quite weakly as the cross section of the ring, namely the number
of channels, increases.
This is the feature expected in the diffusive regime.
On the other hand, when the interaction is taken into account,
the current begins to depend on the cross section.
The current is suppressed for the one-dimensional case, but
is enhanced for the three dimensional case,
the largest sized system being the strongest.

In order to understand the origin of the enhancement,
we have divided current into three parts, namely into the single-particle
term $I_{\rm s}$, the Hartree term $I_{\rm H}$ and the Fock term $I_{\rm F}$,
which are defined as follows, respectively:
$$
I_s(\phi) = - (\partial/\partial\phi)\langle -\sum_{i \ne j}
(t_{i,j}e^{i\theta_{ij}}c_i^\dagger c_j + h.c.)
 + \sum_i \varepsilon_i
c_i^\dagger c_i \rangle,
$$
$$
I_H(\phi) = -(\partial/\partial\phi) \sum_{i \ne j}
V(a/2r_{ij})\langle c_i^\dagger c_i\rangle
\langle c_j^\dagger c_j\rangle,
$$
and
$$
I_F(\phi) = (\partial/\partial\phi) \sum_{i \ne j}
V(a/2r_{ij})\langle c_i^\dagger c_j\rangle
 \langle c_j^\dagger c_i\rangle.
$$
The root mean square of these contributions for the one-dimensional
system and the largest sample,
$6\times 6\times 20$ sites, are shown in Figs.3(a) and (b), respectively.
We see that in the one-dimensional case the contribution from the
interaction terms are smaller than the single electron term, $I_{\rm S}$,
which decreases as $V$ increases.
On the other hand
both the single-particle term and the Hartree term are enhanced
considerably by the Coulomb interaction for the three-dimensional system,
the Fock term remaining quite small.
The former two contribution has different signs,
and the strong enhancement is almost canceled in the total current.

\section{Discussion}
\label{sec:disc}

In the present work we have shown for the first time that the Coulomb
interaction can enhance the persistent current.
It is true that the enhancement is not large enough to be comparable with
the experiment, where the persistent current of the order of pure
one-dimensional ring, $I_0=ev_{\rm F}/L$, has been observed.
The size of the current
 amounts to $I_1 = 4\pi/N_{\ell} \simeq 0.628$, if we use $v_{\rm F}
=2|t|a\hbar$, and $N_{\ell}=20$.
However, the cross section dependence of the enhancement is encouraging,
since the actual sample has width of 90 nm, which is much larger than
the lattice constant of the Au, 0.3 nm.

Then the next question will be whether the strength of the Coulomb interaction
is appropriate or not.
This  is a difficult question,
since size of our system is still too small
compared to the experiment, where about $10^{8}$ atoms are involved.
Here we only comment that if our system is actually made of Au atoms,
$t$ is of the order of 1eV, and bare Coulomb energy between one
lattice constant is larger than $t$: $V = e^2/4\pi \epsilon a
\simeq 5eV > t$.
So even if we consider the optical dielectric constant
$\epsilon(\infty)$
by ion core, $V$ and $t$ are of the same order.
We have found the strongest enhancement when $V$ is a little smaller than $t$.
In the actual system the interaction is screened by electrons,
but such screening effect is included in our self-consistent
solution.
We need the flux dependence of the total ground state energy
which includes the energy associated with the screening.
Therefore use of the long-range Coulomb interaction is needed and
justified.

Our treatment is not exact as we use the Hartree-Fock approximation.
We have justified it by comparison with the results of exact diagonalization
for small one-dimensional systems.
Actually for one-dimensional system the suppression of the current is
stronger in the Hartree-Fock approximation than the exact method.
Thus it is possible that the enhancement
in the 3-d case is stronger in the exact treatment.
However,  more investigation will be needed in order for us to truly
justify and assess our method.

Now
the final question will be the mechanism of the enhancement.
One possible explanation is the screening of the random potential.
The electron at the Fermi level feels screened site-energy which is
flattened, so the current
is enhanced.
In the case of one-dimensional system the screening is also effective,
but electrons are not allowed to pass through each other, thus
the current is suppressed by localization of some of electrons.
When the cross section of the ring is small enough,
this blockade is still effective.
However, Figs.3(a) and (b)  suggest another scenario.
It is seen that $I_{\rm s}$ and $I_{\rm H}$ are enhanced by the
Coulomb interaction considerably for the tree-dimensional system.
Similar behavior is also seen for systems with cross section
larger or equal to $3\times 3$ sites,
but not seen for systems with smaller cross sections.
The large value of the Hartree contribution means that the direct Coulomb
energy changes considerably as the magnetic flux changes.
This is caused by the deformation of the electron wave function as the
flux changes.
The deformation of the wave function does not be done freely.
The Coulomb energy suppresses it for the system
to approach to the charge neutrality.
Thus both the single particle part and the Hartree part are enhanced,
but they nearly cancel each other.
What left of this cancellation brings the enhanced total current.
The finite size of the cross section makes the Coulomb energy larger,
and also gives more freedom for the wave function deform as the flux
changes.
This makes the qualitative difference between the one-dimensional system,
Fig.3(a), and three-dimensional system, Fig.3(b).
Figure 3(b) shows that the ground state is quite different from that of the
non-interacting system,
where we get quite small persistent current.
This explains why the theories using the diagrammatic
expansion\cite{ae,schmid,g2,alt} could not get the enhancement:
Since the ground state has changed considerably from that without
the Coulomb interaction, perturbational
calculation cannot give correct answer.

In conclusion we have shown for the first time that the persistent
current is enhanced by the
long-range Coulomb interaction for three-dimensional rings.
We have clarified that a ring with finite cross section
is necessary for the
enhancement.

\section*{Acknowledgement}

One of the authors (D. Y.) thanks Aspen Center for Physics
where part of this work was done.
The numerical calculation was done by HITAC S-3800/480 at the
Computer Center of the University of Tokyo.


%
%

\begin{figure}
\caption{Fourier component of the persistent current with period
$\phi_0$ is plotted as a function of the strength of the Coulomb
interaction $V/t$ for the randomness $W/t=4.0$.
The five different symbols,
circle, two kinds of triangles, square, and diamond indicate
five different realization of the randomness.
The absolute values of the current are plotted with the open
(closed) symbols indicating that the value is positive (negative).}
\label{fig:v-dep}
\end{figure}

\begin{figure}
\caption{Root-mean-square of the Fourier component of
the persistent current with period $\phi_0$, $I_{\rm typ}$,
is plotted as a function of the strength of the Coulomb
interaction $V/t$ for the randomness $W/t=4.$
Results of six different sized systems are plotted.
The system sizes considered are
 $1\times 1\times 20$ sites with 8 electrons,
 $2\times 2\times 20$ sites with 32 electrons,
 $3\times 3\times 20$ sites with 72 electrons,
 $4\times 4\times 20$ sites with 128 electrons,
 $5\times 5\times 20$ sites with 200 electrons, and
 $6\times 6\times 20$ sites with 300 electrons,
which are shown by open circles, closed circles, open triangles,
closed triangles, open squares, and closed squares, respectively.
}
\label{fig:cross}
\end{figure}

\begin{figure}
\caption{Root-mean-square of the Fourier component of
the three contributions to the total current for the systems with
(a) $1\times 1\times 20$ sites with   8 electrons
and (b)
 $6\times 6\times 20$ sites with 300 electrons.
The single-particle contribution $I_{\rm s}$, the Hartree contribution
$I_{\rm H}$, and
the Fock contribution $I_{\rm F}$ are shown by circles, triangles and squares,
respectively.
It should be noticed that the vertical scale is expanded 10 times in (a).}
\label{fig:comp1}
\end{figure}

\end{document}